# Three-dimensional graphene networks modified with acetylenic linkages for high-performance optoelectronics and Li-ion battery anode material

Chengyong Zhong[1,*] Wenxia Zhang[2], Guangqian Ding[3], Junjie He[1]

[1] Institute for Advanced Study, Chengdu University, Chengdu 610106, China

[2] College of Optoelectronic Engineering, Chongqing University of Posts and Telecommunications, Chongqing 400065, China

[3] School of Science, Chongqing University of Posts and Telecommunications, Chongqing 400065, China

[*] Corresponding Author:

zhongcy90@gmail.com (C. Zhong)

# Abstract

Searching for three-dimensional (3D) semiconducting carbon allotropes with proper bandgaps and excellent optoelectronic properties is always the chasing goal for the new emerging all-carbon optoelectronics. On the other side, 3D carbon materials have also been recognized as promising anode materials superior to commercialized graphite in Li-ion batteries (LIBs). Here, using first-principles calculations, we propose two novel 3D carbon allotropes through acetylenic linkages modification of two structurally intimately correlated 3D carbon structures ― carbon kagome lattice (CKL) and interpenetrated graphene network (IGN). The modified CKL is a truly direct-gap semiconductor and possibly possesses the strongest optical transition coefficient amongst of all semiconducting carbon allotropes. The suitable bandgap and small effective masses also imply it can be a good electron transport material (ETM) for perovskite solar cells. As for the modified IGN, it is a topological nodal-line semimetal and shows greatly enhanced specific capacity as anode materials in LIBs comparing to that of IGN. Our work not only find two new 3D carbon phases with fabulous physical and chemical properties for high-performance optoelectronics and Li-ion anode material, we also offer a fresh view to create various carbon structures with versatile properties.

# 1. Introduction

The potential energy landscape of carbon allotropes is extremely complex, which originate from its unique valence electron configuration $2s^22p^2$, leading to its remarkable hybridization ability[1]. Although naturally existed carbon phases are graphite and diamond but doesn't suggests other metastable carbon structures cannot be synthesized or realized. With the development of experimental techniques, there are more and more novel carbon phases have been reported; for instance, the lately synthesized carbon honeycomb[2, 3] and superhard carbon phase[4] by deposition of vacuum-sublimated graphite and compressing graphite at ambient temperature, respectively. To our surprise, one carbon allotrope (T-Carbon) whose formation energy is far higher than that of ground-state graphite was initially theoretically predicted[5] and then produced in recent experiment by picosecond laser irradiating carbon nanotubes in methanol[6]. These studies continually inspire people to explore novel carbon structures [7-10].

The exploration of novel carbon structures does not just because of people's scientific curiosity about how many carbon allotropes can be found or be realized in laboratories, as archived in Samara Carbon Allotrope Database [11], but also in that carbon materials have amount of important potential applications in different areas. In the field of condensed matter physics, the successful exfoliation of graphene[12] and the companying Dirac fermion [13, 14] has triggered huge attentions on topological physics in carbon[15-21] and non-carbon materials[22-24]. For semiconductor physics, people never stop to pursue the possibility of using carbon-based electronics to replace silicon-based electronics[25, 26]. However, diamond is a wide indirect-gap semiconductor unsuitable for electronic and optoelectronic devices. Therefore, there are highly demands to seek superior semiconducting carbon allotropes. The recently realized T-carbon is indeed a 3D direct-gap semiconductor, being considered as a good ETM for perovskite solar cells[27]. Nevertheless, up to now, high-performance direct-gap 3D carbon semiconductors are still rarely reported.

Otherwise, carbon materials have also been recognized as promising anode materials in LIBs because of their unique advantages, such as abundant in resources, nontoxicity and good Li intercalation and de-intercalation reversibility[28-30]. Recently, one type of 3D graphene networks with topological semimetal features have attracted a lot of attentions as anode materials in LIBs for their intrinsic high conductivity and well-organized ion transport channels[31-36]. For instance, the predicated topological nodal semimetal IGN is claimed as an anode material in LIBs with low diffusion barrier, small volumetric expansion and the theoretical specific capacity is the same as that of graphite anode (372 mA h/g)[33]. Motivated by those recent works on 3D topological semimetal graphene networks for LIBs, one may naturally ask: is it possible find one way to preserve the advantages of 3D graphene networks and meantime to improve the performance for being anode materials in LIBs?

In this work, using first-principles calculations, we propose two novel carbon structures through modification of the dipole-allowed direct-gap semiconductor ― CKL[37], and topological nodal semimetal IGN[15] with acetylenic linkages. To modified CKL, we find that optical transitions between band edge states induced by $p$-orbital frustration are hugely enhanced since there are more $p$ orbital ingredients contributed by acetylenic linkages, making its optical absorption is much better than CKL, GaN, ZnO, and T-Carbon, and maybe the strongest amongst of all semiconducting carbon allotropes; moreover, its bandgap and carrier effective masses are effectively modulated, implying a great potential to be ETM for perovskite solar cells and maybe outperforms over the recently realized T-Carbon. As for the electronic properties of modified IGN, the topological semimetal features are perfectly preserved. Furthermore, with the aid of two simple tight-binding models, we elucidate the underlying orbital interactions in these two intimate carbon structures. Besides, the incorporation of acetylenic linkages in IGN is a practical answer to the question raised above, which the inserted acetylenic linkages make the modified IGN has more space to accommodate Li ions, leading the specific capacity as anode material in LIBs improves about 33.3%. Thus, we not only find two excellent carbon allotropes for high-performance optoelectronics and Li-ion anode material, the approach used here to effectively improves carbon material's diverse properties through acetylenic linkages modification could also give more insights on their various applications in different fields.

## 2. Computational Methods

We performed first-principles calculations within the density functional theory formalism as implemented in VASP code[38, 39]. The potential of the core electrons and the exchange-correlation interaction between the valence electrons are described, respectively, by the projector augmented wave and the generalized gradient approximation (GGA) of Perdew Burke-Ernzerhof (PBE)[40]. The kinetic energy cutoff of 500 eV was employed. The atomic positions were optimized using the conjugate gradient method, the energy convergence value between two consecutive steps was chosen as $10^{-5}$ eV. A maximum force of $10^{-2}$ eV/Å was allowed on each atom. The Brillouin zone (BZ) was sampled with enough *k*-points to make sure the structures were fully relaxed. Phonon calculations were calculated using force-constants method, and the dynamical matrixes were computed using the finite differences method in a 2×2×2 supercell to eliminate errors in the low frequency modes. The force constants and phonon spectra were obtained from the Phonopy package[41].

## 3. Results and Discussion

Considering IGN can be viewed as one of bonds in the triangles of CKL broken (the skeleton of CKL/IGN is shown in Figure 1 with gray balls), we first study CKL under acetylenic linkages modification. As presented in Figs. 1 (a, c), with symmetrical consideration, the C-C bonds (inter-bonds) between two triangles are substituted by acetylenic units (yne-bonds, the blue atoms in Figs. 1a, c), a novel 3D carbon allotrope is obtained, coined as carbon-yne kagome lattice (CYKL) here. Similar process can be performed on intra-bonds of each triangle only or both of inter- and intra-bonds of two triangles simultaneously (see Fig. S1); however, their electronic properties are ruined totally and have no any features (see Fig. S1), comparing with that of CYKL, thus, we wouldn't discuss their results further. The introduction of acetylenic linkages enlarges the lattice parameter *a* (*c*) from 4.46 Å (2.53 Å) to 6.99 Å (6.85 Å) in contrast to its mother CKL (see Table 1). When breaking the left (right) bond of the left (right) triangle in the unit cell of CYKL, another 3D carbon allotrope is obtained (see Figs. 1b, d). Since the presence of acetylenic triple bonds, we call it as interpenetrated graphyne network (IGYN). IGYN, comparing to IGN (see Table 1), the lattice parameter $a$ ($b$/$c$) expands from 5.84 Å (6.39 Å /2.47 Å) to 7.94 Å (11.11 Å /6.89 Å). The inserted acetylenic linkages greatly inflates the volume, the mass density of CYKL (IGYN) shrinks from 2.75 g/cm$^3$ (2.60 g/cm$^3$) to 1.24 g/cm$^3$ (1.18 g/cm$^3$), as given in Table 1, in which we tabulate some structural information of other carbon allotropes for comparison. Small mass densities imply their frameworks have more space to accommodate other atoms or molecules, which is very beneficial for catalysis, gas adsorption and metal atom storage *etc*. Meanwhile, before and after inserting acetylenic linkages, the hybridized types and bond lengths of carbon bonds are also changed, as listed in Table S1, together with diamond, Y-Carbon, T-Carbon, and TY-Carbon, of which Y-Carbon and TY-Carbon are acetylenic linkages modifications of diamond and T-Carbon[42] respectively, for comparison.

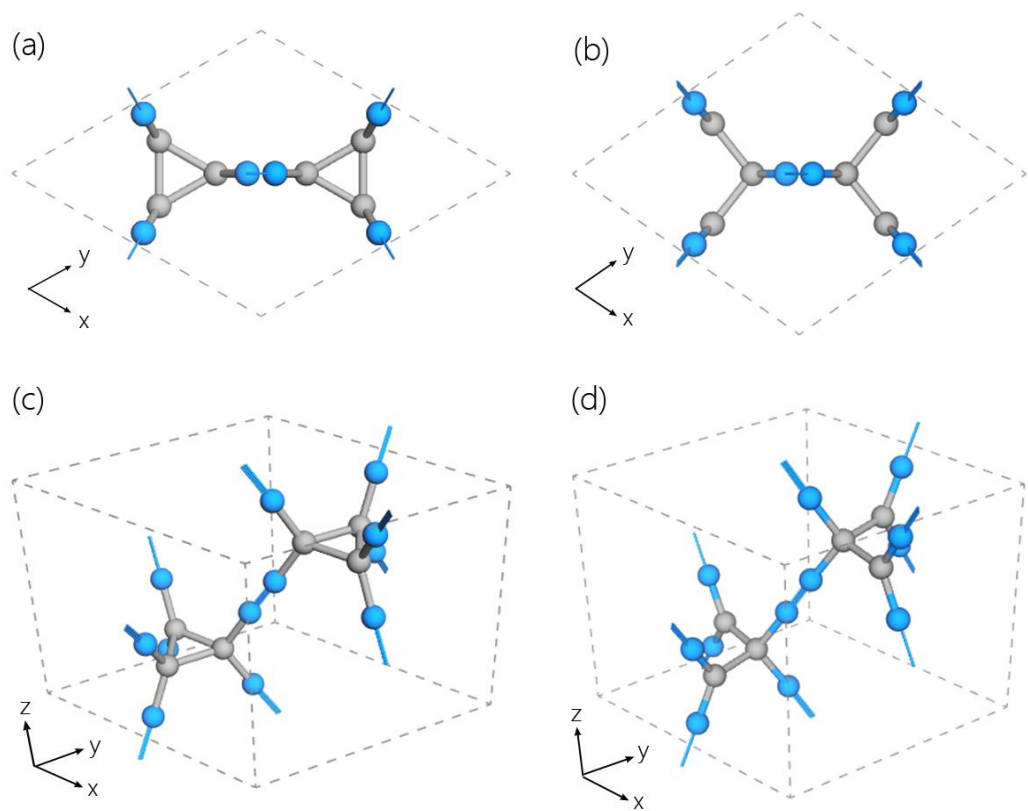

Figure 1. (a, b) The top and (c, d) the perspective views of CYKL, IGYN, respectively. The gray atoms stand for the skeleton of CKL, IGN, the blue atoms are acetylenic linkages.

Table 1. Crystal system, space group (SG), lattice parameters, mass density ρ (g/cm$^3$), bandgap E$_g$ (eV), and total energy E$_t$ (eV/atom) of diamond, Y-Carbon, CKL, CYKL IGN, IGYN, T-Carbon, TY-Carbon, h-Carbon and R16 obtained from our work and other works are given for comparison. The values of bandgap in the parentheses are the results calculated with HSE06.

| system | SG | Method | Lattice parameters | | | ρ | E$_g$ | E$_t$ |
|---|---|---|---|---|---|---|---|---|
| | | | a | b | c | | | |
| Diamond | FD$\bar{3}$M | Our work | 3.56 | | | 3.55 | 4.12 | -9.09 |
| | | GGA[43] | 3.57 | | | 3.50 | (5.34) | -9.09 |
| | | Exp [44] | 3.57 | | | 3.52 | 5.47 | |
| Y-Carbon | | Our work | 9.64 | | | 0.89 | 4.66 | -8.07 |
| | | GGA[42] | 9.64 | | | 0.89 | 4.66 | -8.07 |
| CKL | F6$_3$/MMC | Our work | 4.46 | | 2.53 | 2.75 | 2.26 (3.27) | -8.81 |
| | | GGA[17] | 4.46 | | 2.53 | 2.75 | 2.35 (3.43) | -8.81 |
| CYKL | | Our work | 6.99 | | 6.85 | 1.24 | 1.81(2.71) | -8.24 |
| IGN | CMCM | Our work | 5.84 | 6.39 | 2.47 | 2.6 | Semimetal | -8.98 |
| | | GGA[45] | 5.90 | 6.28 | 2.46 | | Semimetal | -8.85 |
| IGYN | | Our work | 7.94 | 11.11 | 6.89 | 1.18 | Semimetal | -8.29 |
| T-Carbon | FD$\bar{3}$M | Our work | 7.52 | | | 1.50 | 1.47 (2.18) | -7.91 |
| | | GGA[27] | 7.52 | | | 1.50 | (2.27) | -7.92 |
| | | Exp[6] | 7.80 | | | | | |
| TY-Carbon | | Our work | 13.46 | | | 0.53 | 1.47 | -8.03 |
| | | GGA[42] | 13.46 | | | 0.53 | 1.47 | -8.03 |
| h-Carbon | F6$_3$/MCM | Our work | 9.37 | | 4.24 | 2.23 | 1.57 | -9.06 |
| | | GGA[46] | 9.27 | | 4.19 | 2.30 | 1.61 (2.20) | |
| R16 | R$\bar{3}$c | Our work | 6.49 | | 7.76 | 3.38 | 2.99 | -8.58 |
| | | GGA[43] | 6.40 | | 7.66 | 3.39 | (4.23) | -8.59 |

To examine the stability of CYKL and IGYN, we first calculate their total energy as function of the volume per atom, the so-called equation of state (EOS, see Fig. 2a). For comparison, the calculations are also conducted for diamond, Y-Carbon, CKL, IGN, T-Carbon and TY-Carbon. The introduction of acetylenic linkages makes the total energies of the modified structures are higher than that of their respective mothers (see Fig. 2a and Table 1) except of TY-Carbon. The total energies of CYKL and IGYN are lower than that of the experimentally realized T-Carbon 0.33 eV per atom and 0.38 eV per atom, respectively, implying that CYKL and IGYN also have good possibilities to be realized in experiments. Then, we assess their dynamic stability by calculating their phonon spectra (see Figs. 2b, c). One observes that there are no any negative modes throughout the entire BZ, demonstrating they are dynamically stable.

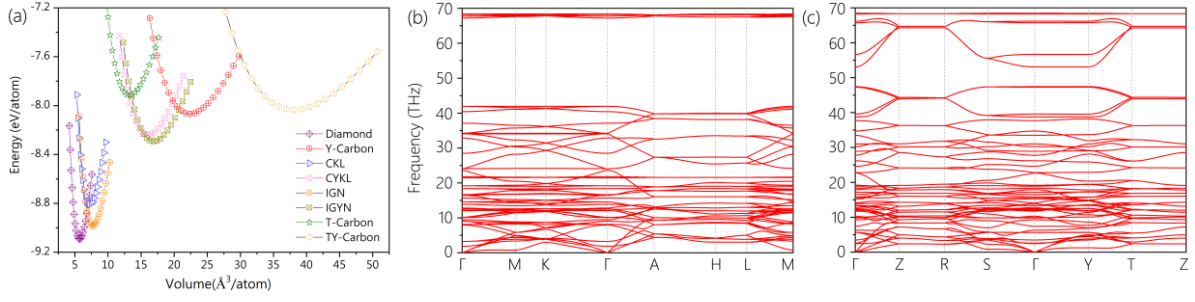

Figure 2. (a) The EOS for diamond, Y-Carbon, CKL, CYKL, IGN, IGYN, T-Carbon and TY-Carbon. (b, c) The phonon spectra for CKYL and IGYN, respectively.

Now we investigate the electronic and optoelectronic properties of CYKL and IGYN. Their band structures calculated from first-principles calculations at PBE level are plotted in Figure 3 and Figure 4. To semiconducting CYKL, we also check its band structure by the more accurate hybrid functional calculation of HSE06[47], which demonstrate the dispersion of each band is almost identical except of there are systematic up/down shift for conduction/valence bands (see Fig. S2b). Therefore, without special statement, we will focus on the PBE results in the following.

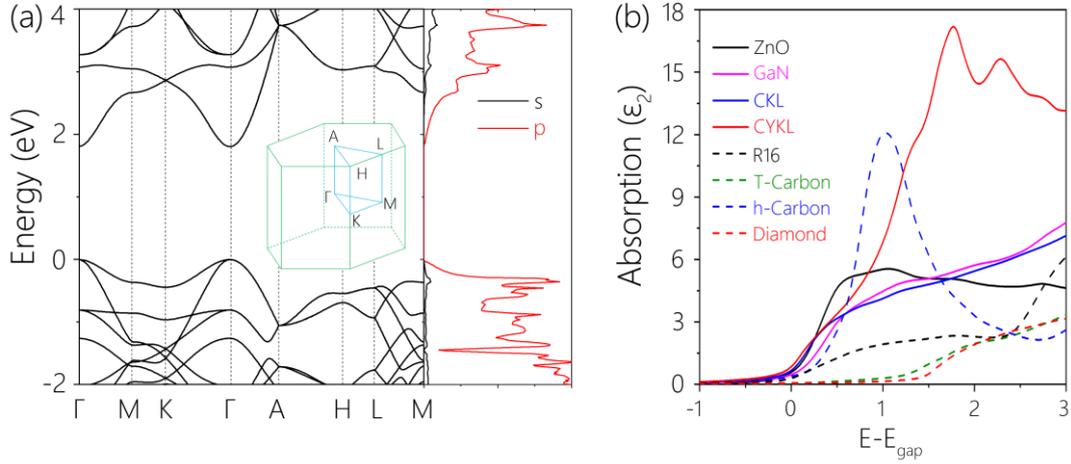

Figure 3. (a) The band structure (left) and PDOS (right) for CYKL, the insert shows the first BZ and the high symmetry $k$-point path. (b) The optical absorption coefficient as function of E - $E_{gap}$ for ZnO, GaN, CKL, CYKL, R16, T-Carbon, h-Carbon and diamond, where $E_{gap}$ is the PBE band gap.

One of key features of CKL is it's a diploe-allowed direct-gap semiconductor induced by $p$ orbitals frustration. One observes that the inserted acetylenic linkages of CYKL do not alter the direct-gap semiconducting characteristic comparing to that of CKL (see Fig. 3a and Figs. S2a, b), except the PBE (HSE06) bandgap diminish from 2.26 eV (3.27 eV) to 1.81 eV (2.71 eV), as listed in Table 1. The projected density of states (PDOS, see Fig. 3a) and the band edge states of CYKL (see Figs. S3a, c) confirm the band edges of conduction/valence bands are mainly contributed from $p$ orbitals. To our surprise, the calculated optical absorption is much stronger than that of ZnO, GaN, CKL, the recently realized direct-gap semiconductor — T-Carbon, direct-gap semiconducting R16/h-Carbon[43, 46], and diamond (see Fig. 3b). We believe the strong optical transitions between the conduction and valence bands are also the results of $p$ orbitals frustration, which still exists in the triangles (see Figs. S3 a, c), meanwhile the $p$ orbitals transitions is hugely enhanced since there are more $p$ orbitals contribution from the two unpaired $p$ orbitals in the acetylenic triple bonds.

Table 2. The electron and hole effective masses for CKL, CYKL and T-Carbon. The notation used here are the same as Ref [37]. The effective masses are given in unit of the free electron mass.

| system | $m_e^{\parallel}$ | $m_{lh}^{\parallel}$ | $m_{hh}^{\parallel}$ | $m_e^{\perp}$ | $m_h^{\perp}$ |
|---|---|---|---|---|---|
| CKL | 0.16 | -0.12 | -0.51 | 1.23 | -0.91 |
| CYKL | 0.34 | -0.33 | -1.09 | 0.31 | -0.61 |
| T-Carbon | 3.98 | -0.18 | -0.52 | 3.54 | -0.27 |

To the experimentally realized T-Carbon, Sun *et.al*[27] claim it is a good ETM for perovskite solar cells because of its moderate direct bandgap and small effective carrier mass, even superior to conventional ETMs such as TiO$_2$, ZnO and SnO$_2$. Through comprehensive comparison of the electron (hole) effective masses of CKL, CYKL and T-Carbon (see Table 2 and Fig. S2), we find the electron effective masses $m_e^{\parallel}$ ($m_e^{\perp}$) of CKL 0.16 (1.23) and CYKL 0.34 (0.31) are much smaller than that of T-Carbon 3.98 (3.54) and the hole effective masses are also comparable to that of T-Carbon (see Table 2). On the other hand, as an efficient ETM, it is supposed to be colorless, because the optical absorption of ETM should not affect the light-harvesting of the perovskites in the visible region (350 nm ~ 770 nm)[48]. Therefore, the ideal bandgap of ETM should be around or larger than 3 eV. The HSE06 bandgaps of CKL and CYKL are 3.27 eV and 2.71 eV. Thus, from the discussion above, we can argue that CKL and CYKL maybe also good candidates as ETMs, even have better performance than T-Carbon.

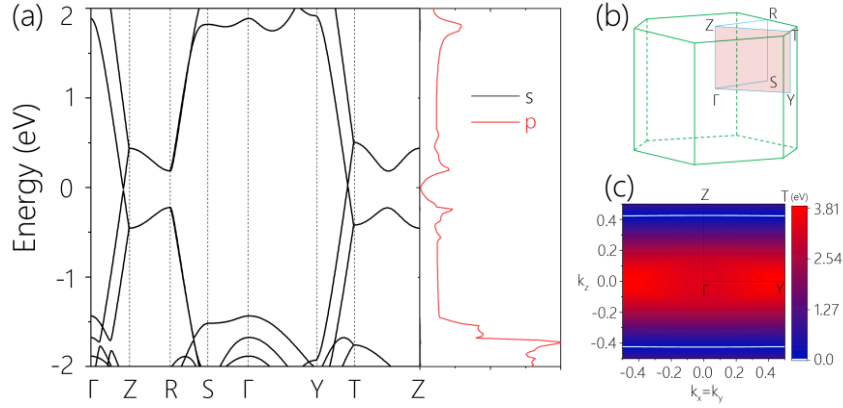

Figure 4. (a) The band structure (left), PDOS (right), (b) the first BZ and the high symmetry $k$-point path of IGYN, the red shadow part is used to calculate energy difference between valence band and conduction band, as shown in (c), where the white lines indicate the location of nodal lines.

The band structure of IGYN is shown in Figure 4. One observes that the conduction band and valence band linearly cross two Dirac points at Fermi level along Γ-Z and Y-T. After carefully scanning the whole reciprocal space, in fact, there are two lines consisting of Dirac points in the BZ (see the bright lines in Fig. 4b), forming the so-called nodal line as happened in its mother IGN. Therefore, the acetylenic linkages do not destroy the nodal lines, but more interesting, make the nodal lines more straight than that of IGN[15].

Next, in order to elucidate the underlying orbital interactions in CYKL and IGYN, we build their tight-binding (TB) models here. From the perspective of TB theory, the inserted acetylenic linkages in CYKL and IGYN do not change the symmetry of CKL and IGN, because they just play the role of transferring the hopping interactions. Thus, when we construct the TB models, the acetylenic linkages can be dropped safely. Based on the PDOS and orbitals analysis above, we propose the CKL-TB and IGN-TB models as shown in Figure 5(a, c) with consideration of single orbital in each site. The TB Hamiltonian reads:

$$H = \sum_{<i,j>} t_{i,j} c_i^\dagger c_j \qquad (1)$$

where $c_i^\dagger$ and $c_j$ represent the creation and the annihilation operators, respectively. $t_{i,j}$ is the hopping energy between orbitals at sites i and j. In detail, we use $t_0$ and $t_1$ to describe the nearest-neighbor hopping energies of intra- and inter-triangles; $t_2$ and $t_3$ represent the high-order next-neighbor hopping energies (Fig. 5 a). As for IGN-TB model, $t_0$ and $t_1$ stand for the hopping interactions between and along two zigzag chains (see Fig. 5c). Figs. 5(b, d) show the TB band structures for the CYKL and IGYN. By comparing Fig. 3/Fig. 4 with Fig. 5, one observes that the band structure of TB theory fits to that of first-principles calculations around Fermi level very well.

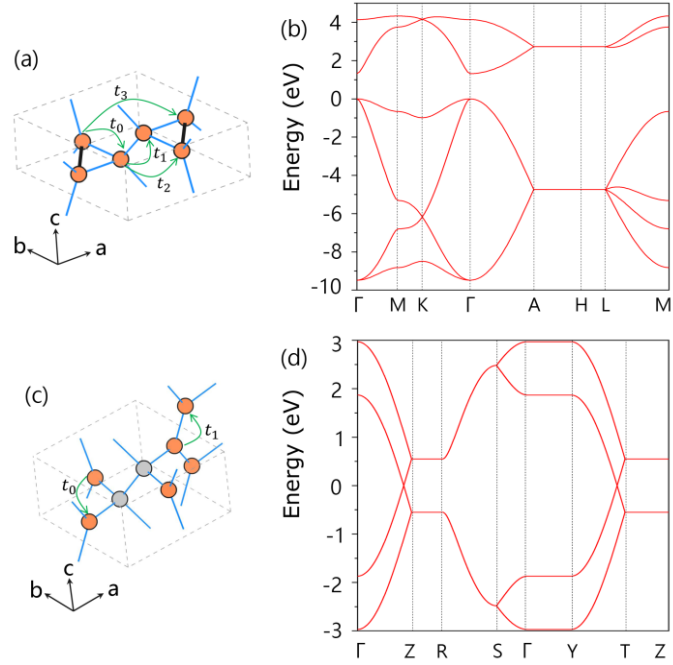

Figure 5. (a) and (c) indicate the hopping parameters considered in the TB models for CKL and IGN, respectively. The orange balls are the effective sites used for TB modeling, the gray balls in IGN-TB model are not participate in the formation of band structures near fermi level considering they are $sp^3$ atoms. The bonds colored with black in (a) are broken for IGN-TB model, leading the main interactions change a lot. (b) and (d) are the TB electronic band structures correspondingly. The values of fitted hopping parameters are presented in Table 3.

Table 3. The tight-binding parameters for the CKL- and IGN-TB models. The corresponding hopping processes are depicted in Figure 5 (a, c).

|     | $t_0$ | $t_1$ | $t_2$ | $t_3$ |
| --- | --- | --- | --- | --- |
| CKL | 2.49 | 1.82 | -0.20 | -0.17 |
| IGN | 0.55 | 1.21 |      |      |

Liu *et.al*[33] reported that IGN as a topological nodal line semimetal porous carbon could be a promising new anode material in LIBs for its intrinsic conductivity and regular transport channels. After inserting acetylenic linkages, the properties of topological nodal line semimetal preserved and there are more storage sites and space to accommodate Li-ion simultaneously, thus, it's highly desire to explore the possibility of being anode material in LIBs for IGYN.

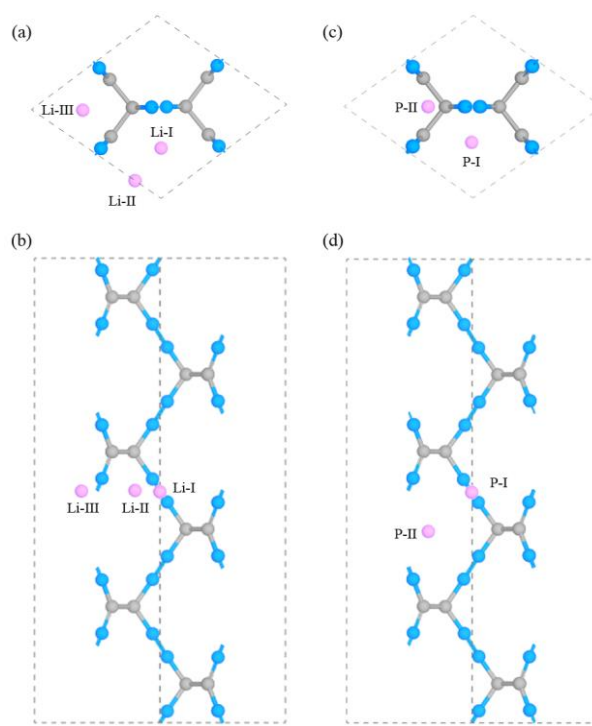

Figure 6. (a) Top and (b) side views of the possible Li-ion adsorption sites in the 1×1×3 supercell of IGYN. The pink, gray, and blue balls mean the Li, $sp^3$- ($sp^2$-), and $sp$ - hybridized carbon atoms, respectively. (c) Top and (d) side views of the fully optimized locations, with the initial configurations indicated in (a). The site Li-I relaxes to P-I, the sites Li-II and Li-III relax to P-II.

We first investigate the favorable intercalated sites for single Li atom. To avoid the interaction between Li atoms, we adopt a 1×1×3 supercell. Three possible locations are considered, marked as Li-I, Li-II and Li-III (see Figs. 6a, b). Li-I (Li-II) is the site nearby central (peripheral) zigzag carbyne chain, its height locates at the central of the acetylenic bond. Li-III is the site between two peripheral zigzag carbyne chains, its height is same as Li-I (Li-II). After fully relaxed, Li-I almost holds its initial position (marked as P-I, see Figs. 6c, d), however Li-II and Li-III relax to the same position (marked as P-II, see Figs. 6c, d). The adsorption energies of Li atom for P-I and P-II are -0.75 eV and -1.13 eV calculating with the adsorption energy formula:

$$E_{ad} = (E_{Li_xC} - E_C - xE_{Li})/x \qquad (2)$$

where $E_{Li_xC}$ / $E_C$ is the total energy of IGYN with/without intercalated Li atom, $E_{Li}$ is the atomic energy of single Li in the bulk, $x$ is the number of Li atoms under different Li ion concentration.

We next calculate the energy barrier for Li ion diffusion. Here, we consider two paths going through P-I and P-II and paralleling to the *c*-axis. The calculated diffusion barriers along paths I and II are 0.3 eV and 0.45 eV (see Fig. S4a), calculating with the climbing-image nudge elastic band (CI-NEB) method[49]. The small energy barriers are comparable to that of graphite (see Table 4)[50], indicating the Li ions can easily transport along the channels.

Table 4. The diffusion barrier, specific capacity and volume change of IGN, IGYN, graphite.

| System | Diffusion barrier (eV) | Specific capacity (mAh/g) | Volume change |
|---|---|---|---|
| IGN [33] | 0.04 | 372 | ~3.2% |
| IGYN | 0.3 - 0.45 | 496 | ~2.1% |
| Graphite [50] | 0.22 - 0.4 | 372 | ~10% |

In the end, the maximum Li capacity of IGYN is determined through increasing the Li atoms until the full Li-intercalated configuration is reached (see Fig. S4b). According to our calculations, one unit cell of IGYN can host four Li atoms with volumetric expansion of about 2.1 % (the absorption energy of Li ion intercalated in IGYN under different concentration is shown in Fig S5a). The determined chemical formula can be described as LiC$_{4.5}$, corresponding to the Li storage capacity is 496 mA h /g, which is larger than that of graphite / IGN (372 mA h /g), as listed in Table 4.

Another one of the key factors to characterize anode material in LIBs is the open circuit voltage (OCV). Commonly, the OCV of anodes should be low to get a maximum capacity for LIBs[51]. It has been reported that the anode may not work when the OCV is above 2 V, thus the OCV of anode materials should be between 0 ~ 2 V[52]. The difference in total energies before and after Li intercalation is used to determine the OCV, which can be approximately calculated with formula[53]:

$$E_{OCV} = -(E_{Li_xC} - E_C - xE_{Li})/xe \qquad (3)$$

where the meaning of each term is same as that of formula (2), $e$ is the charge of electron. The OCV

of maximum capacity of IGYN ($LiC_{4.5}$) is 0.67 V (the OCV of Li ion intercalated in IGYN under different concentration is shown in Fig S5b), which a bit larger than that of graphite (0.3 V)[54] but smaller than that of $TiO_2$ (1.5 – 1.8 V)[55, 56] and black phosphorene (1.8 – 2.9 V)[57]. In general, the voltage will decrease to 0 V with the increase of Li concentration (in the charge process), hence suitable and higher intercalation potential is benefit to accommodate more Li atoms. Therefore, IGYN is still an ideal anode material for LIBs. Otherwise, the OCV possibly can be decreased by defects or doping, which are waiting for more researches.

Before concluding, we will give a few remarks. For better utilization of carbon-based electronics and optoelectronics in the future, people never cease to search appropriate semiconducting carbon materials with suitable direct bandgap, high carrier mobility or strong optical absorption *etc*. Although, there are plenty of methods to open and tune the bandgap in graphene, however, there are rare reports about 3D direct-gap carbon semiconductors in the experiments, except of the recently realized T-Carbon. In 2D carbon materials, one of effective way to obtain direct-gap semiconductor is by inserting acetylenic linkages in graphene network, such as the way to synthesize graphdiyne[58], which has a direct bandgap and high carrier mobility[59]. As for 3D carbon materials, the scheme of incorporating acetylenic linkages in 3D graphene networks is still an adoptable approach to acquire desirable 3D semiconducting carbon materials, as manifesting in the CYKL. The 3D direct-gap carbon allotropes are already hardly found, letting alone the truly dipole-allowed ones. To our knowledge, CYKL maybe the most promising 3D semiconducting carbon material for it possibly possesses the best and strongest optical absorption in all 3D carbon allotropes.

As anode material in LIBs, the most commercialized graphite exhibits large volume change (about 10%) due to the weak vander Waals force between different layers, leading to the poor cycling performance at high rate, and its specific capacity is also limited by the weak interaction between Li ion and graphite[30]. In contrast, the advantages of IGYNs as anode materials in LIBs can be summarized as: (1) IGYNs can be viewed as different graphyne nanosheets covalently connecting by small graphyne segments, which can greatly minimize volume change during the process of charging/discharging. (2) The unpaired valence electrons in the acetylenic linkages provide more interaction sites between Li ions and the host, which can furtherly improve the host's storage capacity. (3) As topological carbon semimetal, 3D IGYNs possess extremely well-organized ion transport channels with intrinsic high electronic conductivity, which is very beneficial for high rate cycling performance. Therefore, we do believe 3D IGYNs have great potential application as anode materials in LIBs.

Carbon phases can possess numerous forms with different hybridized bonding types[1]. The introduction of acetylenic linkages makes the hybridized bonding types change from $sp^3$ ($sp^2+sp^3$) in CKL (IGN) to $sp+sp^3$ ($sp+sp^2+sp^3$) in CYKL (IGYN), which enrich not only the family of carbon allotropes but also their physical and chemical properties. Even though carbon allotropes can have multiple hybridized bonds in one structure, but contain simultaneously all $sp$, $sp^2$ and $sp^3$ hybridized bonds with intriguing physical and chemical properties are rarely reported[60]. Very recently, You *et al.*[61] proposed a novel carbon allotrope named Carboneyane is a topological nodal line semimetal with $sp+sp^2+sp^3$ chemical bonds, which just similar to that of IGYN here. Therefore, we envision the insertion of acetylenic triple bonds into carbon phases maybe can spawn more fascinating materials with distinctive properties[62].

In addition, although we discuss semiconducting properties of CYKL and the potential of IGYN as Li-ion anode material here, but doesn't hints they are constrained to those areas only. The large specific surface area and porous makes them can be also used in some other fields, such as catalysis, energy storage and molecular sieves[9].

To facilitate the experimental identification of CYKL and IGYN, we simulate the X-ray diffraction (XRD) patterns for CYKL/IGYN, and compare it with diamond as well as with the experimental data from detonation soot of TNT, diesel soil[63], and chimney soot[64], as plotted in Figure 7. The experimental measured XRD spectra usually contain plenty of crystalline phases. One can observes that the diffraction peaks (200) in CYKL and (020) in IGYN are very close to the peaks around 2θ = 27° and 30° in the detonation soot and chimney soot, which suggests CYKL and IGYN are likely the candidates for the crystalline phases in the detonation soot and chimney soot.

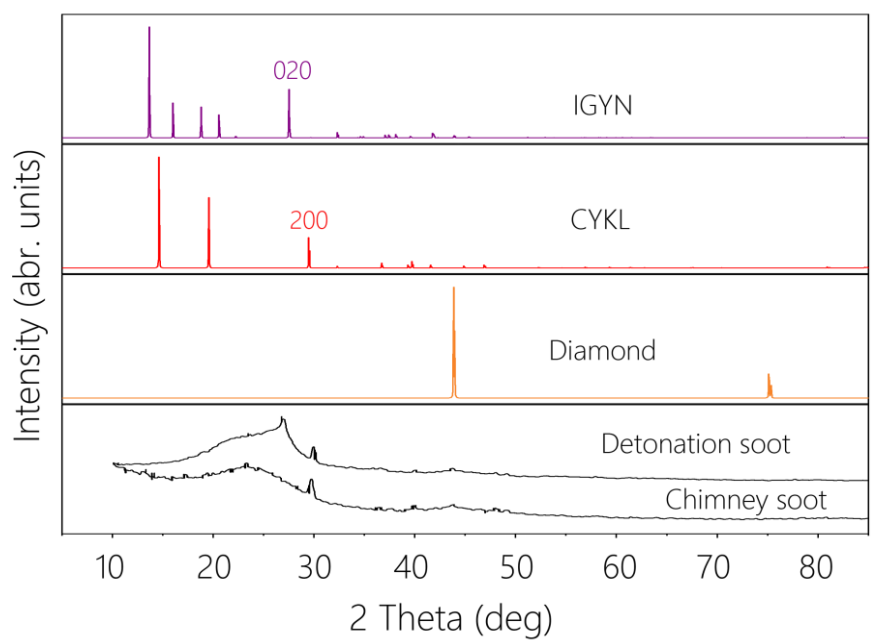

Figure 7. Simulated XRD patterns for CYKL and IGYN, compared with diamond as well as experimental XRD patterns for detonation soot of TNT, diesel oil [63]and chimney soot[64].

## 4. Conclusion

In conclusion, based on first-principles calculations, we discover two novel 3D carbon phases, the CYKL and IGYN, through inserting acetylenic linkages into CKL and IGN, respectively. With the incorporated acetylenic units, CYKL exhibits possibly the strongest optical absorption amongst of all carbon allotropes, stemming from the strength of $p$ orbitals frustration enhanced by the unpaired $p$ orbitals in acetylenic bonds. On the other hand, the features of topological nodal-line semimetal is preserved in IGYN. Moreover, the inserted acetylenic linkages make IGYN showing great improved specific capacity as anode material in LIBs since the additional intercalated space. Additionally, the volume change is also small (about 2.1 %) and the ion diffusion barriers comparable to that of graphite. Our work not only reveals two novel carbon allotropes with interesting properties, but also provides an alternative route for tailoring optoelectronic properties in 3D semiconducting carbon structures and improving the specific capacity of anode materials in LIBs

# Acknowledgements

The authors thank Junping Hu for his valuable discussion. This work is partially supported by the National Natural Science Foundation of China (Grant Nos. 11804039, 11804040, 11804041). W. Z. acknowledges support by the Natural Science Foundation of Chongqing (Grant No: cstc2018jcyjAX0547). Particularly, C. Z. wants to thank Y.N. Zhang for her encouragement, support and care over past days, 9966?